\documentclass{article}
\usepackage{ijcai24}

\usepackage{times}
\usepackage{soul}
\usepackage{url}
\usepackage[hidelinks]{hyperref}
\usepackage[utf8]{inputenc}
\usepackage[small]{caption}
\usepackage{graphicx}
\usepackage{amsmath}
\usepackage{amsthm}
\usepackage{booktabs}
\usepackage{algorithm}
\usepackage{algorithmic}
\usepackage[switch]{lineno}
\usepackage{graphicx}
\usepackage{booktabs} 
\usepackage{bm}
\usepackage{amsmath}
\usepackage{graphicx,amssymb}
\usepackage{multirow}
\usepackage{threeparttable}
\usepackage{verbatim}
\usepackage[misc]{ifsym}
\usepackage{hyperref}
\usepackage{graphicx}
\usepackage{microtype}
\usepackage{graphicx}
\usepackage{subfig}
\usepackage{booktabs} 
\usepackage{amssymb}
\usepackage[normalem]{ulem}
\usepackage{dsfont}
\usepackage{multirow}
\usepackage{makecell}
\newcommand{\TheName}{GS$^2$P}
\title{Generative Pre-trained Ranking Model with Over-parameterization at Web-Scale (Extended Abstract)\thanks{This work was accepted by the Sister Conference Track of IJCAI 2024.}
}

\author{
Yuchen Li$^{1,2}$\and
Haoyi Xiong$^2$\and
Linghe Kong$^1$\and
Jiang Bian$^2$\and
Shuaiqiang Wang$^2$\and \\
Guihai Chen$^1$\And
Dawei Yin$^2$
\\
\affiliations
$^1$Shanghai Jiao Tong University, China~~~
$^2$Baidu Inc., China\\
\emails
\{yuchenli, linghe.kong, gchen\}@sjtu.edu.cn,
haoyi.xiong.fr@ieee.org,\\
\{jiangbian03, shqiang.wang\}@gmail.com,
yindawei@acm.org
}
\begin{document}

\maketitle

\begin{abstract}
\emph{Learning to rank} (LTR) is widely employed in web searches to prioritize pertinent webpages from retrieved content based on input queries. However, traditional LTR models encounter two principal obstacles that lead to suboptimal performance: (1) the lack of well-annotated query-webpage pairs with ranking scores covering a diverse range of search query popularities, which hampers their ability to address queries across the popularity spectrum, and (2) inadequately trained models that fail to induce generalized representations for LTR, resulting in overfitting.
To address these challenges, we propose a \emph{\uline{G}enerative \uline{S}emi-\uline{S}upervised \uline{P}re-trained} (\TheName{}) LTR model. We conduct extensive offline experiments on both a publicly available dataset and a real-world dataset collected from a large-scale search engine. Furthermore, we deploy \TheName{} in a large-scale web search engine with realistic traffic, where we observe significant improvements in the real-world application.
\end{abstract}

\section{Introduction}
The booming increase of internet users and web content surges the demands on web search. In the current digital epoch, large-scale search engines manage an impressive archive of trillions of webpages, providing service to hundreds of millions of active users daily while handling billions of queries~\cite{xiong2024search,liao2024towards,chen2024temporalmed,lyu2022joint1}. The search procedure commences with a user query, often a text string, necessitating keyword or phrase extraction to comprehend user attempting~\cite{DBLP:conf/coling/ZhaoWL10,li2023mhrr,chen2024hotvcom,chen2024recent,lyu2024semantic}. Post identification of keywords, search engines evaluate the relation between the query and webpages, subsequently retrieving highly relevant ones from their vast databases~\cite{huang2021learning,yu2018adaptively,lyu2022joint}. These webpages are then sorted based on content attributes and click-through rates, positioning the most relevant ones on top of the result~\cite{li2023mpgraf,xiong2024towards,chen2023xmqas,chen2024emotionqueen,chen2024xmecap,lyu2023semantic}.

The optimization of the user experience, achieved by catering to information needs, largely depends on the effective sorting of retrieved content. In this realm, Learning to Rank (LTR) becomes instrumental, requiring a considerable amount of query-webpage pairings with relevancy scores for effective supervised LTR~\cite{li2023s2phere,DBLP:journals/corr/QinL13,li2023ltrgcn,lyu2020movement,peng2024graphrare,wang2024soft}. Nevertheless, the commonplace scarcity of well-described, query-webpage pairings often compels semi-supervised LTR, harnessing both labeled and unlabeled samples for the process~\cite{DBLP:conf/cikm/SzummerY11,DBLP:journals/www/ZhangHL16,zhu2023pushing,peng2023clgt}.
Recent years have seen the integration of deep models in LTR, aimed at end-to-end ranking loss minimization~\cite{DBLP:conf/icpr/LiL0R20,DBLP:conf/www/WangSCJLHC21,li2022meta,DBLP:journals/csur/YangY23,chen2024talk,chen2022grow}. However, these models occasionally falter in learning generalizable representations from structural data due to limited or noisy supervision, sometimes resulting in performance that is weaker compared to statistical learners~\cite{DBLP:conf/sigir/BruchZBN19,lyu2024rethinking,wang2024multi,chen2023mapo}. Further discussion on this subject can be found in a comprehensive review available in a recent scholarly work~\cite{DBLP:journals/ml/Werner22,chen2023can,chen2023hadamard,chen2023hallucination,chen2024drAcademy}.

\begin{figure*}[t]
	 	\centering
 		\includegraphics[width=0.85\textwidth]{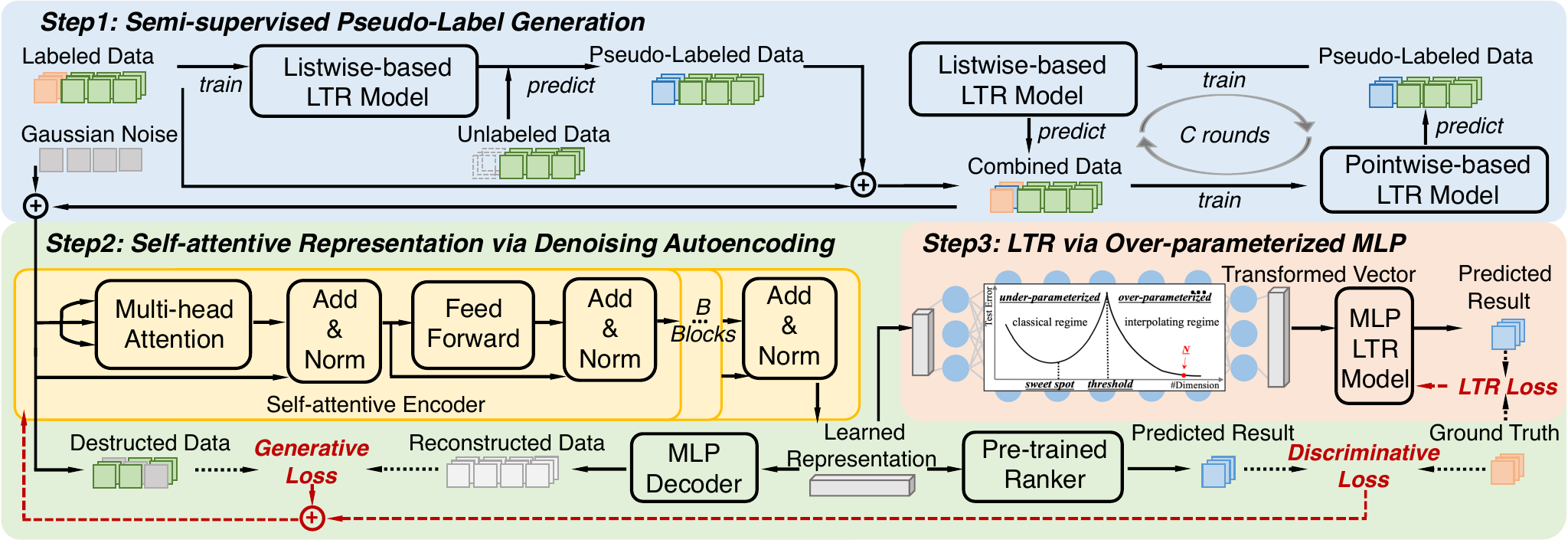}
 		\caption{The framework of \TheName{}.}
 		\label{pipeline}
\end{figure*}   

In order to tackle the above issues, we propose a Generative Semi-Supervised Pre-trained LTR (\TheName{}) model.
The proposed \TheName{} first generates high-quality pseudo labels for every unlabeled query-webpage pair through co-training of multiple/diverse LTR models based on various ranking losses, then learns generalizable representations with a self-attentive network using both generative loss and discriminative loss. Finally, given the generalizable representations of query-webpage pairs, by incorporating an MLP-based ranker with Random Fourier Features (RFF), \TheName{} pushes LTR models into so-called interpolating regime~\cite{belkin2021fit,song_going_2023,chen2022beyond,chen2024explanations,cai2023resource} and obtains superb performance improvement. 
To demonstrate the effectiveness of \TheName{}, we conduct comprehensive experiments on a publicly available LTR dataset~\cite{DBLP:journals/corr/QinL13,chen2024dolarge} and a real-world dataset collected from a large-scale search engine. We also deploy \TheName{} at the search engine and evaluate the proposed model using online A/B tests in comparison with the online legacy system.

\section{Methodology}
\subsection{Preliminaries}
Given a set of search queries $\mathcal{Q}=\{q_1,~q_2,$ $\dots \}$ and all archived webpages $\mathcal{W}=\{w_1,w_2,\dots\}$, for each query $q_i\in\mathcal{Q}$, the search engine retrieves a set of relevant webpages denoted as $W_i=\{w^i_1,w^i_2,\dots\}\subset\mathcal{W}$. 
After annotating, each query $q_i$ is assigned with a set of relevance scores $\boldsymbol{y}_i=\{y^i_1,y^i_2,\dots\}$. 
In this work, we follow the settings in~\cite{DBLP:journals/corr/QinL13,li2023coltr} and scale the relevance score from 0 to 4  to represent levels of relevance, which represents whether the webpage w.r.t. the query is bad (0), fair (1), good (2), excellent (3) or perfect (4).
We denote a set of query-webpage pairs with relevance score annotations as $\mathcal{T}^L=\{(q_1,W_1,\boldsymbol{y}_1),~(q_2,W_2,\boldsymbol{y}_2),\dots \}$.
The core problem of semi-supervised LTR is to leverage unlabeled pairs, i.e., $\mathcal{T}^U=\{(q'_1,W'_1),~(q'_2,W'_2),\dots\}\subset\mathcal{Q}$ and $\vert\mathcal{T}^U\vert\gg \vert\mathcal{T}^L\vert$, in the training process.

\subsection{Semi-supervised Pseudo-Label Generation}
Given the overall set of queries $\mathcal{Q}$ and the set of all webpages $\mathcal{W}$, \TheName{} first obtains every possible query-webpage pair from both datasets, denoted as $(q_i, w_i^j)$ for $\forall q_i\in \mathcal{Q}$ and $\forall w_i^j\in W_i\subset\mathcal{W}$, i.e.,  the $j^{th}$ webpage retrieved for the $i^{th}$ query. For each query-webpage pair $(q_i, w_i^j)$, \TheName{} further extracts an $m$-dimensional feature vector $\boldsymbol{x}_{i,j}$ representing the features of the $j^{th}$ webpage under the $i^{th}$ query.
Then, the labeled and unlabeled sets of feature vectors can be presented as $\mathcal{D}^L=\{(\boldsymbol{x}_{i,j},\boldsymbol{y}^i_j)\vert \forall (q_i,W_i,\boldsymbol{y})\in\mathcal{T}^L~\text{and}~\forall w_j^i\in W_i\}$ and $\mathcal{D}^U=\{\boldsymbol{x}_{i,j}\vert \forall (q_i,W_i)\in\mathcal{T}^U\}$.
Inspired by~\cite{li2023coltr}, \TheName{} leverages a semi-supervised learning LTR manner to generate high-quality pseudo labels for unlabeled samples.

\subsection{Self-attentive Representation Learning via Denoising Autoencoding}
\textbf{Denoised Self-attentive Autoencoder.}
Given an $m$-dimensional feature vector $\tilde{\boldsymbol{x}}_{i,j}$ of a query-webpage pair $(\tilde{\boldsymbol{x}}_{i,j},\boldsymbol{y}^i_j)$ in combined data, \TheName{} aims to utilize a self-attentive encoder to learn a generalizable representation $\boldsymbol{z}_{i,j}$.
Specifically, given a vector $\tilde{\boldsymbol{x}}_{i,j}$ generated from \emph{Semi-supervised Pseudo-Label Generation}, \TheName{} (1) passes it through a fully-connected layer and produces a hidden representation.
Then, \TheName{} (2) feeds the hidden representation into a self-attentive autoencoder, which consists of $B$ encoder blocks of Transformer~\cite{DBLP:conf/nips/VaswaniSPUJGKP17}.
In particular, each encoder block incorporates a multi-head attention layer and a feed-forward layer, both followed by layer normalization.
Eventually, \TheName{} (3) generates the learned representation $\boldsymbol{z}_{i,j}$ from the last encoder block.
For each original feature vector $\tilde{\boldsymbol{x}}_{i,j}$, the whole training process can be formulated as $\boldsymbol{z}_{i,j} = f_{\tilde{\theta}}(\tilde{\boldsymbol{x}}_{i,j})$,
where $\tilde{\theta}$ is the set of parameters of the self-attentive encoder.

Given the learned representation $\boldsymbol{z}_{i,j}$, \TheName{} leverages an MLP-based decoder for the reconstruction task.
Specifically, for each representation $\boldsymbol{z}_{i,j}$ produced from the self-attentive autoencoder, \TheName{} uses the MLP-based decoder to map $\boldsymbol{z}_{i,j}$ to a generalizable representation $\boldsymbol{z}^{{\prime}}_{i,j}$, which has the same dimension with the original feature vector.
The whole training process can be formulated as $\boldsymbol{z}^{{\prime}}_{i,j} = g_{\theta^{\prime}}(\boldsymbol{z}_{i,j})$,
where the $\theta^{\prime}$ is the set of parameters of the MLP-based decoder.
Finally, \TheName{} jointly optimizes the parameter sets $\tilde{\theta}$ and $\theta^{\prime}$ to minimize the generative loss as $\mathcal{L}_\text{G} = 
    \frac{1}{\vert \mathcal{Q}\vert} \frac{1}{\vert W_i\vert}\sum_{i=1}^{\vert \mathcal{Q}\vert}\sum_{j=1}^{\vert W_i\vert} \ell_\text{G}\left(\tilde{\boldsymbol{x}}_{i,j}, \boldsymbol{z}^{{\prime}}_{i,j}\right)$,
where $\ell_\text{G}$ is the squared error, which could be presented as $\ell_\text{G}\left(\tilde{\boldsymbol{x}}_{i,j}, \boldsymbol{z}^{{\prime}}_{i,j}\right)=\|\tilde{\boldsymbol{x}}_{i,j}-\boldsymbol{z}^{{\prime}}_{i,j}\|^{2}$.

\textbf{Pre-trained Ranker.}
Given the learned vector $\boldsymbol{z}_{i,j}$ generated from \emph{Denoised Self-attentive Autoencoder}, \TheName{} leverages a fully-connected layer to obtain predicted scores $\boldsymbol{r}_{i,j}$ as $\boldsymbol{r}_{i,j} = k_{\theta}(\boldsymbol{z}_{i,j})$,
where $\theta$ is the set of discriminative parameters of \emph{Pre-trained Ranker}.
Against the ground truth, \TheName{} utilizes the discriminative loss function $\mathcal{L}_\text{D}$ to compute the loss of ranking prediction as $\mathcal{L}_\text{D} = 
    \frac{1}{\vert \mathcal{Q}\vert} \frac{1}{\vert W_i\vert}\sum_{i=1}^{\vert \mathcal{Q}\vert}\sum_{j=1}^{\vert W_i\vert} \ell_\text{D}\left(\boldsymbol{y}^i_j, \boldsymbol{r}_{i,j}\right)$,
where $\ell_\text{D}$ is denoted as the standard LTR loss function.
Then, \TheName{} jointly optimizes the discriminative loss $\mathcal{L}_\text{D}$ and the generative loss $\mathcal{L}_\text{G}$ to accomplish both discriminative (LTR) and generative (denoising autoencoding for reconstruction) tasks simultaneously as $\mathcal{L}_\text{Final} = \alpha\mathcal{L}_\text{D} + \beta\mathcal{L}_\text{G}$,
where $\alpha, \beta \in[0,1]$ are weight coefficients to balance two terms.

\begin{table*}[t]
\centering
 {
 \scriptsize
 \begin{tabular}{l  c c  c c  c c  c c}
 \toprule
\multirow{2}{*}{\textbf{Methods}} &
 \multicolumn{2}{c}{\textbf{5\%}}  & \multicolumn{2}{c}{\textbf{10\%}} &
 \multicolumn{2}{c}{\textbf{15\%}} & \multicolumn{2}{c}{\textbf{20\%}} \\
 \cmidrule(r){2-3}
\cmidrule(r){4-5}
\cmidrule(r){6-7}
\cmidrule(r){8-9}
 & $@{4}$ & $@{10}$ & $@{4}$ & $@{10}$ & $@{4}$ & $@{10}$ & $@{4}$ & $@{10}$ \\
 \midrule
 XGBoost        & 31.76  & 34.10  & 36.72  & 39.12  & 39.93  & 41.01  & 42.60 & 45.84  \\
 LightGBM       & 35.72  & 39.32  & 39.89  & 42.05  & 43.90  & 45.67  & 46.56 & 48.52  \\
\midrule
 RMSE       & 34.82 & 38.02  & 38.75  & 41.95  & 42.97  & 45.65  & 45.75  & 48.86 \\
 RankNet    & 34.06 & 37.43  & 38.12  & 41.32  & 42.24  & 45.08  & 45.01  & 47.89 \\
 LambdaRank & 35.28 & 38.50  & 39.32  & 42.47  & 43.40  & 46.23  & 46.26  & 49.56 \\
 ListNet    & 34.36 & 37.94  & 38.31  & 41.76  & 42.51  & 45.40  & 45.32  & 48.42  \\
 ListMLE    & 33.47 & 36.95  & 37.52  & 40.84  & 41.53  & 44.43  & 44.39  & 47.26  \\
 ApproxNDCG & 33.98 & 37.20  & 37.94  & 41.01  & 42.09  & 44.70  & 44.94  & 47.50 \\
 NeuralNDCG & 35.15 & 38.26  & 39.07  & 42.10  & 43.32  & 45.97  & 46.08  & 49.20  \\
 \midrule
 $\rm{CR_{RMSE}}$       & 36.04  & 38.54  & 39.52  & 42.48  & 43.67  & 46.25  & 46.86 & 49.75 \\
 $\rm{CR_{RankNet}}$    & 35.90  & 38.42  & 39.44  & 42.37  & 43.45  & 45.98  & 46.70 & 49.61 \\
 $\rm{CR_{LambdaRank}}$ & 36.45  & 38.93  & 40.03  & 43.10  & 44.36  & 46.88  & 47.57 & 50.47 \\
 $\rm{CR_{ListNet}}$    & 37.53  & 40.08  & 41.28  & 44.21  & 45.17  & 47.73  & 48.35 & 51.24 \\
 $\rm{CR_{ListMLE}}$    & 35.67  & 38.16  & 39.40  & 42.35  & 43.28  & 45.86  & 46.62 & 49.48  \\
 $\rm{CR_{ApproxNDCG}}$ & 37.93  & 40.41  & 41.47  & 44.32  & 45.53  & 48.03  & 48.81 & 51.69  \\
 $\rm{CR_{NeuralNDCG}}$ & 37.26  & 40.65  & 40.76  & 43.69  & 44.85  & 47.52  & 48.16 & 51.13  \\
 \midrule
 \TheName{}$_{\rm{RMSE}}$     & 39.02 & 40.88 & 41.80 & 44.72 & 45.72 & 48.22 & 48.72 & 51.40  \\
 \TheName{}$_{\rm{RankNet}}$  & 38.15 & 40.42 & 40.03 & 44.21 & 44.93 & 47.85 & 47.85 & 50.98  \\
 \TheName{}$_{\rm{LambdaRank}}$ & 39.47 & 41.43  & 42.17  & 45.20 & 46.07 & 48.89 & 49.15 & 51.97 \\
 \TheName{}$_{\rm{ListNet}}$    & 39.53  & 41.62  & 42.28  & 45.42  & 46.15  & 49.16  & 49.18 & 52.20  \\
 \TheName{}$_{\rm{ListMLE}}$   & 37.66 & 39.87  & 39.80  & 43.70  & 44.52  & 47.28 & 47.41  & 50.24  \\
 \TheName{}$_{\rm{ApproxNDCG}}$    & 39.57  & 41.76 & 42.39 & 45.65  & 46.31  & 49.31  & 49.25  & 52.25 \\
 \TheName{}$_{\rm{NeuralNDCG}}$   & \textbf{39.72} & \textbf{41.97} & \textbf{42.56} & \textbf{45.83} & \textbf{46.38} & \textbf{49.53} & \textbf{49.36} & \textbf{52.47} \\
 \bottomrule
 \end{tabular}
 }
\caption{Results for Web30K on NDCG across diverse labeled data percentages.}
\label{web30k}
\end{table*}

\subsection{LTR via Over-parameterized MLP}
Given the learned representation $\boldsymbol{z}_{i,j}\in\mathcal{R}^n$ generated from \emph{Self-attentive Representation Learning via Denoising Autoencoding}, \TheName{} converts this representation vector into an $N$-dimensional version, represented as $\boldsymbol{h}_{i,j}=\mathbf{h}(\boldsymbol{z}_{i,j})$. This step is implemented using the feature transformation $\mathbf{h}(z)$. 
In this procedure, \TheName{} utilizes a transformation rooted in random Fourier features to execute $\mathbf{h}(z)$~\cite{DBLP:conf/nips/RahimiR07}, thereby mapping the original features of LTR into a higher dimensional feature space.
An important point to consider is that increasing the number of dimensions ($N$) leads to over-parameterization of the LTR model via the addition of more input features. This scenario brings about a feature-wise 'double descent' phenomenon in predicting generalization errors~\cite{belkin2019reconciling,belkin2021fit}. 
\TheName{} sets the optimal value for $N$, stemming from cross-validation performed on the labeled dataset to ensure the best generalization performance. Therefore, incorporating $\boldsymbol{h}_{i,j}$ for every pair of query-webpage paves the path for an over-parameterized LTR model. This advanced model operates in the interpolating regime and is projected to exhibit excellent generalization performance~\cite{belkin2021fit}. 
In this way, \TheName{} transforms $\boldsymbol{z}_{i,j}$ into a high-dimensional vector $\boldsymbol{h}_{i,j}$ and constructs a Ranker (i.e., MLP-based LTR model) for the LTR task with several popular ranking loss functions.

\section{Experiments}
\subsection{Experimental Setup}
\textbf{Datasets.}
We carry out the offline experiments on a standard and publicly available dataset Web30K~\cite{DBLP:journals/corr/QinL13} and a real-world dataset \emph{commerical dataset} collected from Baidu search engine.
Specifically, the commercial Dataset contains 50,000 queries. The dataset is annotated by a group of professionals on the crowdsourcing platform, who assign a score between 0 and 4 to each query-document pair.

\textbf{Metrics.}
To assess the performance of various ranking systems comprehensively, we leverage the following metrics. 
Normalized Discounted Cumulative Gain (NDCG)~\cite{DBLP:journals/sigir/JarvelinK17} is a standard listwise accuracy metric, which has been commonly used in research and industrial community. For our online evaluation, we utilize the Good vs. Same vs. Bad ($\mathrm{GSB}$)~\cite{zhao2011automatically}, which is an online pairwise-based evaluation methodology evaluated by annotators. Considering the confidentiality of commercial information, we only report the difference between the results of \TheName{} and the online \emph{legacy system}.

\begin{table*}[t]
\centering
 {
 \scriptsize
 \begin{tabular}{l  c c  c c  c c  c c}
 \toprule
\multirow{2}{*}{\textbf{Methods}} &
 \multicolumn{2}{c}{\textbf{5\%}}  & \multicolumn{2}{c}{\textbf{10\%}} &
 \multicolumn{2}{c}{\textbf{15\%}} & \multicolumn{2}{c}{\textbf{20\%}} \\
 \cmidrule(r){2-3}
\cmidrule(r){4-5}
\cmidrule(r){6-7}
\cmidrule(r){8-9}
 & $@{4}$ & $@{10}$ & $@{4}$ & $@{10}$ & $@{4}$ & $@{10}$ & $@{4}$ & $@{10}$ \\
 \midrule
 XGBoost        & 48.39  & 52.12  & 52.83  & 56.45  & 56.14  & 60.03  & 58.03 & 62.61  \\
 LightGBM       & 50.48  & 53.50  & 54.13  & 59.04  & 57.00  & 62.14  & 60.47 & 65.82  \\
 \midrule
 RMSE       & 49.73  & 53.42  & 54.13  & 57.86  & 57.43  & 61.34  & 59.42  & 64.76 \\
 RankNet    & 49.32  & 53.07  & 53.76  & 57.37  & 57.08  & 60.92  & 59.17  & 64.25 \\
 LambdaRank & 50.82  & 54.24  & 55.07  & 58.62  & 58.16  & 62.05  & 61.12  & 65.28  \\
 ListNet    & 50.26  & 53.61  & 54.52  & 58.04  & 57.81  & 61.47  & 59.74  & 64.82 \\
 ListMLE    & 48.73  & 52.46  & 53.08  & 56.70  & 56.32  & 60.25  & 58.42  & 63.68  \\
 ApproxNDCG & 49.08  & 52.75  & 53.44  & 57.02  & 56.79  & 60.61  & 58.84  & 64.01 \\
 NeuralNDCG & 50.68  & 53.89  & 54.88  & 58.31  & 58.02  & 61.82  & 61.03  & 64.97  \\
 \midrule
 $\rm{CR_{RMSE}}$       & 50.43  & 53.63  & 54.52  & 58.70  & 56.90  & 61.74  & 60.42 & 65.22 \\
 $\rm{CR_{RankNet}}$    & 50.86  & 54.06  & 54.98  & 58.26  & 57.32  & 61.82  & 60.83 & 65.61 \\
 $\rm{CR_{LambdaRank}}$ & 52.47  & 55.67  & 56.13  & 59.84  & 58.90  & 63.79  & 61.87 & 66.59 \\
 $\rm{CR_{ListNet}}$    & 52.45  & 55.64  & 56.08  & 59.82  & 58.74  & 63.24  & 62.28 & 67.09 \\
 $\rm{CR_{ListMLE}}$    & 51.05  & 54.30  & 54.76  & 58.46  & 57.53  & 62.01  & 61.04 & 65.83  \\
 $\rm{CR_{ApproxNDCG}}$ & 51.92  & 55.08  & 55.68  & 59.40  & 58.42  & 62.87  & 62.00 & 66.75  \\
 $\rm{CR_{NeuralNDCG}}$ & 52.06  & 55.31  & 55.87  & 59.61  & 58.67  & 63.20  & 62.18 & 66.84  \\
 \midrule
 \TheName{}$_{\rm{RMSE}}$ & 52.72  & 55.48  & 55.89  & 59.60  & 58.82  & 63.13  & 61.92  & 66.24 \\
 \TheName{}$_{\rm{RankNet}}$ & 53.13  & 55.93  & 56.20  & 59.92  & 58.94  & 63.41  & 62.28  & 66.67  \\
 \TheName{}$_{\rm{LambdaRank}}$ & 53.67  & 56.72  & 56.90  & 60.76  & 59.58  & 64.19  & 62.95  & 67.65  \\
 \TheName{}$_{\rm{ListNet}}$ & 54.00  & 57.18  & 57.28  & 61.04  & 59.93  & 64.50  & 63.38  & 67.96  \\
 \TheName{}$_{\rm{ListMLE}}$ & 53.41  & 56.24  & 56.51 & 56.51 & 59.20  & 63.72  & 62.50  & 66.88   \\
 \TheName{}$_{\rm{ApproxNDCG}}$ & 54.23  & 57.32 & 57.44  & 61.12 & 60.12  & 64.62  & 63.58  & 68.05 \\
 \TheName{}$_{\rm{NeuralNDCG}}$ & \textbf{54.36 } & \textbf{57.43 } & \textbf{57.62 } & \textbf{61.25 } & \textbf{60.28 }& \textbf{64.76 } & \textbf{63.72 } & \textbf{68.12 }\\
 \bottomrule
 \end{tabular}
 }
\caption{Results for Commercial Dataset on NDCG across diverse labeled data percentages.}
\label{commercial}
\end{table*}

\begin{table}[t]
\centering
{
\scriptsize
\begin{tabular}{l  cc  cc }
\toprule
\multirow{2}{*}{} &
\multicolumn{2}{c}{\bf{\TheName{}$_{\rm{ApproxNDCG}}$}} & \multicolumn{2}{c}{\bf{\TheName{}$_{\rm{NeuralNDCG}}$} }\\
\cmidrule(r){2-3}
\cmidrule(r){4-5}
& Random & Long-Tail & Random & Long-Tail\\
\midrule
$\Delta GSB$ 
& \textbf{+3.00\%} & \textbf{+4.00\%} & \textbf{+5.50\%} & \textbf{+6.50\%}\\
\bottomrule
\end{tabular}
}
\caption{Performance improvements of \TheName{} with ApproxNDCG loss and \TheName{} with NeuralNDCG loss for the online evaluation.}
\label{online1}
\end{table}

\textbf{Loss Functions and Competitor Systems}
In this work, we leverage the following advanced ranking loss functions to evaluate the proposed model comprehensively, such as RMSE, RankNet~\cite{DBLP:conf/icml/BurgesSRLDHH05}, LambdaRank~\cite{DBLP:conf/nips/BurgesRL06},
ListNet~\cite{DBLP:conf/icml/CaoQLTL07},
ListMLE~\cite{DBLP:conf/icml/XiaLWZL08},
ApproxNDCG~\cite{DBLP:journals/ir/QinLL10},
and NeuralNDCG~\cite{DBLP:journals/corr/abs-2102-07831}.

As for the ranking model, we choose the following state-of-the-art ranking models as the competitor for \TheName{}, such as MLP, Context-aware Ranker (CR)~\cite{context}, XGBoost~\cite{xgboost} and LightGBM~\cite{lightgbm}.

\subsection{Offline Experimental Results}
\textbf{Overall Results.}
Table~\ref{web30k} and~\ref{commercial} present the average results for offline evaluation, where \TheName{} is compared with competitors on Web30K and the commercial dataset. 
Intuitively, we could observe \TheName{} outperforms all competitors with different losses under various ratios of labeled data on two datasets.
More specifically, \TheName{} with NeuralNDCG gets 3.60\% and nearly 3.57\% higher NDCG$@4$ and NDCG$@10$ on Web30K dataset, compared with the pointwise-based self-trained MLP model with NeuralNDCG.
On Commercial Dataset, \TheName{} on average obtains nearly 2.84\% and 3.14\% improvement on NDCG$@4$ and NDCG$@10$, when compared with NeuralNDCG.
\TheName{}+NeuralNDCG could gain the most improvement under the less ratio of labeled data on both metrics on two datasets, which demonstrates the effectiveness of \TheName{} under low-resource situations.

\subsection{Online Evaluation}

\begin{figure}[t]
    \centering
    \includegraphics[width=0.46\textwidth]{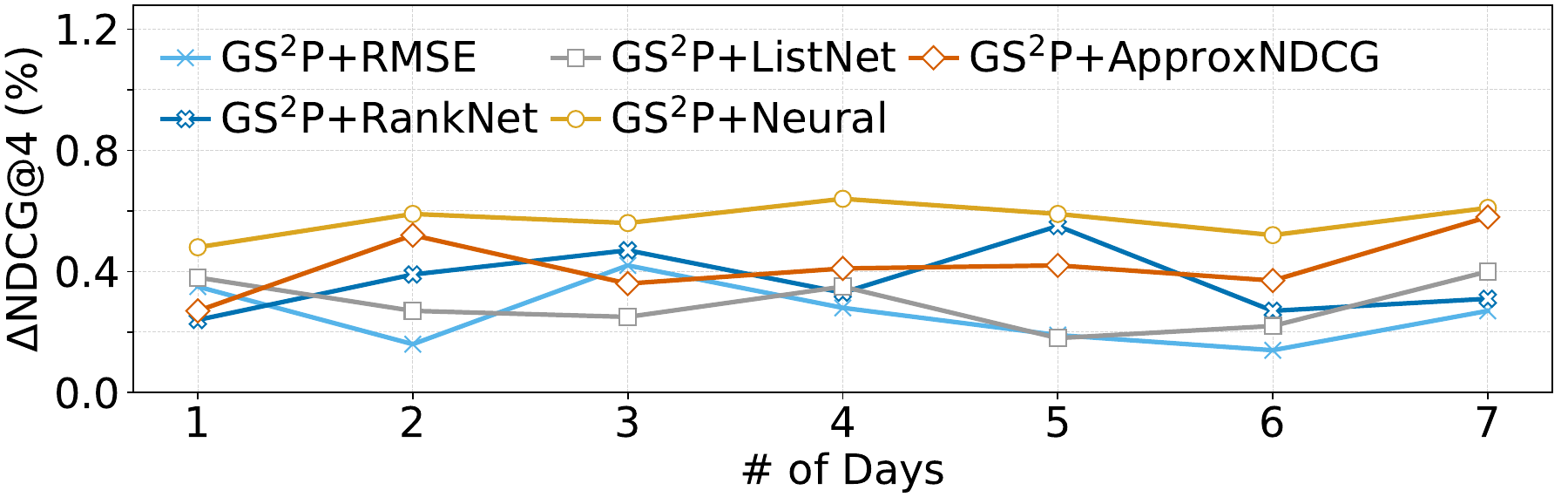}
    \caption{A/B test results of \TheName{} and the legacy system for 7 days (\emph{t}-test with $p < 0.05$ over the baseline). 
    }
    \label{online}
\end{figure}

To comprehensively evaluate our proposed model, we conduct a manual comparison experiment.
Intuitively, manual comparison results are presented in Table~\ref{online1}.
In particular, we observe that our proposed model outperforms the online legacy system by a large margin for random and long-tail (i.e., the search frequency of the query is lower than 10 per week) queries.
Specifically, \TheName{} with NeuralNDCG loss achieves the largest improvement compared with the legacy system with 5.50\% and 6.50\% for random and long-tail queries, respectively. Moreover, \TheName{} with ApproxNDCG loss also improves the performance for random and long-tail queries. 
Figure~\ref{online} illustrates the relative performance between \TheName{} and the base model, expressed via $\Delta NCDG@4$. Logically, \TheName{} shows marked enhancement in performance across all days when compared to the base system, evidencing its practical capability in upgrading the efficacy of a large-scale search engine. Even more impressively, \TheName{} has shown substantial growth on this large-scale platform. A prominent highlight is \TheName{} outperforming the online base model by a significant margin of 0.61\% relative improvement on $\Delta NCDG@4$, a feat achieved by the NeuralNDCG loss-trained model using a nominal 5\% labeled data ratio. 
\TheName{} has showcased consistent performance across both online and offline platforms. 

\section{Related Works}
To enhance the user experience in search results, ranking the retrieved content is a crucial step in which the LTR model plays a significant role. Based on the loss function, LTR models can be categorized into three families: pointwise, pairwise~\cite{DBLP:conf/icml/BurgesSRLDHH05,DBLP:conf/nips/BurgesRL06}, and listwise~\cite{DBLP:conf/icml/CaoQLTL07,DBLP:journals/ir/QinLL10}.
For the pointwise loss, the LTR problem is formulated as a regression task~\cite{zhou2024explainable}.
The pairwise approach converts the problem into binary classification by considering document pairs. 
Besides, the listwise approach treats the entire document list as a single sample and directly optimizes the evaluation metrics~\cite{zhou2024pass,wang2023sst,wang2023st}.

The core principle of data reconstruction involves learning the joint probability distribution of samples from the training data. Variational autoencoder architectures~\cite{DBLP:journals/corr/KingmaW13,DBLP:conf/cvpr/Tran0ZJ17,liang2024model} have been utilized for data reconstruction in prior studies~\cite{liu2024robustifying,lu2024cats}. For instance,~\cite{DBLP:conf/cvpr/Tran0ZJ17} employed a cascaded residual autoencoder inspired by the denoised autoencoder~\cite{DBLP:journals/jmlr/VincentLLBM10,zhou2024portfolio,Weng202404} to estimate residuals and reconstruct corrupted multimodal data sequences. 

The issue of balancing under-parameterization and over-parameterization has received increasing attention from researchers~\cite{belkin2019reconciling,zhou2024application,chen2022cross,xin2024let,wang2024uncertainty}. To address LTR tasks,~\cite{DBLP:conf/cikm/SzummerY11,jin2024agentreview,jin2024mm,jin2024better} proposes a parameter learning-based approach, which has shown that over-parameterization can achieve superior performance on test data under certain conditions in the "interpolation regime" with the double descent curve~\cite{shangguan2021trend,zheng2021makes,liang2021omnilytics}. 
Random Fourier Feature~\cite{DBLP:conf/nips/RahimiR07,jin2022prototypical,ding2024llava,ni2024timeseries,weng2024leveraging,weng2024leveraging,chen2023dual} method utilizes a kernel technique to generate features for most inner product-based models, which has demonstrated significant improvements~
\cite{jiang2024multi,xie2024wildfiregpt,fu2024detecting,wang2023novel}. 

\section{Conclusion}
In this work, we design, implement and deploy a generative semi-supervised pre-trained model \TheName{} on a real-world large-scale search engine to address the problems of LTR under semi-supervised settings.
We substantiate the effectiveness of \TheName{} through comprehensive offline and online analyses, juxtaposed against an extensive lineup of rivals. The offline trials denote a considerable leap in \TheName{}'s performance relative to other baselines. Furthermore, \TheName{} significantly enhances the online ranking efficacy in practical applications, mirroring the positive outcomes observed in the offline experiments.

\section*{Acknowledgments}
This work was initially presented at the 10th IEEE International Conference on Data Science and Advanced Analytics (DSAA) in 2023 and Machine Learning (MLJ) in 2024~\cite{li2024gs2p}.

\bibliographystyle{named}
\bibliography{reference}

\begin{thebibliography}{}

\bibitem[\protect\citeauthoryear{Belkin \bgroup \em et al.\egroup }{2019}]{belkin2019reconciling}
Mikhail Belkin, Daniel Hsu, Siyuan Ma, and Soumik Mandal.
\newblock Reconciling modern machine-learning practice and the classical bias--variance trade-off.
\newblock {\em Proceedings of the National Academy of Sciences}, 116(32):15849--15854, 2019.

\bibitem[\protect\citeauthoryear{Belkin}{2021}]{belkin2021fit}
Mikhail Belkin.
\newblock Fit without fear: remarkable mathematical phenomena of deep learning through the prism of interpolation.
\newblock {\em Acta Numerica}, 30:203--248, 2021.

\bibitem[\protect\citeauthoryear{Bruch \bgroup \em et al.\egroup }{2019}]{DBLP:conf/sigir/BruchZBN19}
Sebastian Bruch, Masrour Zoghi, Michael Bendersky, and Marc Najork.
\newblock Revisiting approximate metric optimization in the age of deep neural networks.
\newblock In {\em Proceedings of the 42nd International {ACM} {SIGIR} Conference on Research and Development in Information Retrieval}, pages 1241--1244, 2019.

\bibitem[\protect\citeauthoryear{Burges \bgroup \em et al.\egroup }{2005}]{DBLP:conf/icml/BurgesSRLDHH05}
Christopher J.~C. Burges, Tal Shaked, Erin Renshaw, Ari Lazier, Matt Deeds, Nicole Hamilton, and Gregory~N. Hullender.
\newblock Learning to rank using gradient descent.
\newblock In {\em Machine Learning, Proceedings of the Twenty-Second International Conference, {ICML}}, pages 89--96, 2005.

\bibitem[\protect\citeauthoryear{Burges \bgroup \em et al.\egroup }{2006}]{DBLP:conf/nips/BurgesRL06}
Christopher J.~C. Burges, Robert Ragno, and Quoc~Viet Le.
\newblock Learning to rank with nonsmooth cost functions.
\newblock In {\em Advances in Neural Information Processing Systems 19, Proceedings of the Twentieth Annual Conference on Neural Information Processing Systems}, pages 193--200, 2006.

\bibitem[\protect\citeauthoryear{Cai \bgroup \em et al.\egroup }{2023}]{cai2023resource}
Mingxin Cai, Yutong Liu, Linghe Kong, Guihai Chen, Liang Liu, Meikang Qiu, and Shahid Mumtaz.
\newblock Resource critical flow monitoring in software-defined networks.
\newblock {\em IEEE/ACM Transactions on Networking}, 32(1):396--410, 2023.

\bibitem[\protect\citeauthoryear{Cao \bgroup \em et al.\egroup }{2007}]{DBLP:conf/icml/CaoQLTL07}
Zhe Cao, Tao Qin, Tie{-}Yan Liu, Ming{-}Feng Tsai, and Hang Li.
\newblock Learning to rank: from pairwise approach to listwise approach.
\newblock In {\em Machine Learning, Proceedings of the Twenty-Fourth International Conference}, pages 129--136, 2007.

\bibitem[\protect\citeauthoryear{Chen and Guestrin}{2016}]{xgboost}
Tianqi Chen and Carlos Guestrin.
\newblock Xgboost: A scalable tree boosting system.
\newblock In {\em Proceedings of the 22nd acm sigkdd international conference on knowledge discovery and data mining}, pages 785--794, 2016.

\bibitem[\protect\citeauthoryear{Chen and Xiao}{2024}]{chen2024recent}
Yuyan Chen and Yanghua Xiao.
\newblock Recent advancement of emotion cognition in large language models.
\newblock 2024.

\bibitem[\protect\citeauthoryear{Chen \bgroup \em et al.\egroup }{2022a}]{chen2022beyond}
Jiamin Chen, Xuhong Li, Lei Yu, Dejing Dou, and Haoyi Xiong.
\newblock Beyond intuition: Rethinking token attributions inside transformers.
\newblock {\em Transactions on Machine Learning Research}, 2022.

\bibitem[\protect\citeauthoryear{Chen \bgroup \em et al.\egroup }{2022b}]{chen2022grow}
Yuyan Chen, Yanghua Xiao, and Bang Liu.
\newblock Grow-and-clip: Informative-yet-concise evidence distillation for answer explanation.
\newblock In {\em 2022 IEEE 38th International Conference on Data Engineering (ICDE)}, pages 741--754. IEEE, 2022.

\bibitem[\protect\citeauthoryear{Chen \bgroup \em et al.\egroup }{2022c}]{chen2022cross}
Zheng Chen, Yulun Zhang, Jinjin Gu, Linghe Kong, Xin Yuan, et~al.
\newblock Cross aggregation transformer for image restoration.
\newblock {\em Advances in Neural Information Processing Systems}, 35:25478--25490, 2022.

\bibitem[\protect\citeauthoryear{Chen \bgroup \em et al.\egroup }{2023a}]{chen2023hadamard}
Yuyan Chen, Qiang Fu, Ge~Fan, Lun Du, Jian-Guang Lou, Shi Han, Dongmei Zhang, Zhixu Li, and Yanghua Xiao.
\newblock Hadamard adapter: An extreme parameter-efficient adapter tuning method for pre-trained language models.
\newblock In {\em Proceedings of the 32nd ACM International Conference on Information and Knowledge Management}, pages 276--285, 2023.

\bibitem[\protect\citeauthoryear{Chen \bgroup \em et al.\egroup }{2023b}]{chen2023hallucination}
Yuyan Chen, Qiang Fu, Yichen Yuan, Zhihao Wen, Ge~Fan, Dayiheng Liu, Dongmei Zhang, Zhixu Li, and Yanghua Xiao.
\newblock Hallucination detection: Robustly discerning reliable answers in large language models.
\newblock In {\em Proceedings of the 32nd ACM International Conference on Information and Knowledge Management}, pages 245--255, 2023.

\bibitem[\protect\citeauthoryear{Chen \bgroup \em et al.\egroup }{2023c}]{chen2023can}
Yuyan Chen, Zhixu Li, Jiaqing Liang, Yanghua Xiao, Bang Liu, and Yunwen Chen.
\newblock Can pre-trained language models understand chinese humor?
\newblock In {\em Proceedings of the Sixteenth ACM International Conference on Web Search and Data Mining}, pages 465--480, 2023.

\bibitem[\protect\citeauthoryear{Chen \bgroup \em et al.\egroup }{2023d}]{chen2023mapo}
Yuyan Chen, Zhihao Wen, Ge~Fan, Zhengyu Chen, Wei Wu, Dayiheng Liu, Zhixu Li, Bang Liu, and Yanghua Xiao.
\newblock Mapo: Boosting large language model performance with model-adaptive prompt optimization.
\newblock In {\em Findings of the Association for Computational Linguistics: EMNLP 2023}, pages 3279--3304, 2023.

\bibitem[\protect\citeauthoryear{Chen \bgroup \em et al.\egroup }{2023e}]{chen2023xmqas}
Yuyan Chen, Yanghua Xiao, Zhixu Li, and Bang Liu.
\newblock Xmqas: Constructing complex-modified question-answering dataset for robust question understanding.
\newblock {\em IEEE Transactions on Knowledge and Data Engineering}, 2023.

\bibitem[\protect\citeauthoryear{Chen \bgroup \em et al.\egroup }{2023f}]{chen2023dual}
Zheng Chen, Yulun Zhang, Jinjin Gu, Linghe Kong, Xiaokang Yang, and Fisher Yu.
\newblock Dual aggregation transformer for image super-resolution.
\newblock In {\em Proceedings of the IEEE/CVF international conference on computer vision}, pages 12312--12321, 2023.

\bibitem[\protect\citeauthoryear{Chen \bgroup \em et al.\egroup }{2024a}]{chen2024explanations}
Jiamin Chen, Xuhong Li, Yanwu Xu, Mengnan Du, and Haoyi Xiong.
\newblock Explanations of classifiers enhance medical image segmentation via end-to-end pre-training.
\newblock {\em arXiv preprint arXiv:2401.08469}, 2024.

\bibitem[\protect\citeauthoryear{Chen \bgroup \em et al.\egroup }{2024b}]{chen2024dolarge}
Yuyan Chen, Yueze Li, Songzhou Yan, Sijia Liu, Jiaqing Liang, and Yanghua Xiao.
\newblock Do large language models have problem-solving capability under incomplete information scenarios?
\newblock In {\em Proceedings of the 62nd Annual Meeting of the Association for Computational Linguistics}, 2024.

\bibitem[\protect\citeauthoryear{Chen \bgroup \em et al.\egroup }{2024c}]{chen2024hotvcom}
Yuyan Chen, Songzhou Yan, Qingpei Guo, Jiyuan Jia, Zhixu Li, and Yanghua Xiao.
\newblock Hotvcom: Generating buzzworthy comments for videos.
\newblock In {\em Proceedings of the 62nd Annual Meeting of the Association for Computational Linguistics}, 2024.

\bibitem[\protect\citeauthoryear{Chen \bgroup \em et al.\egroup }{2024d}]{chen2024drAcademy}
Yuyan Chen, Songzhou Yan, Panjun Liu, and Yanghua Xiao.
\newblock Dr.academy: A benchmark for evaluating questioning capability in education for large language models.
\newblock In {\em Proceedings of the 62nd Annual Meeting of the Association for Computational Linguistics}, 2024.

\bibitem[\protect\citeauthoryear{Chen \bgroup \em et al.\egroup }{2024e}]{chen2024emotionqueen}
Yuyan Chen, Songzhou Yan, Sijia Liu, Yueze Li, and Yanghua Xiao.
\newblock Emotionqueen: A benchmark for evaluating empathy of large language models.
\newblock In {\em Proceedings of the 62nd Annual Meeting of the Association for Computational Linguistics}, 2024.

\bibitem[\protect\citeauthoryear{Chen \bgroup \em et al.\egroup }{2024f}]{chen2024xmecap}
Yuyan Chen, Songzhou Yan, Zhihong Zhu, Zhixu Li, and Yanghua Xiao.
\newblock Xmecap: Meme caption generation with sub-image adaptability.
\newblock In {\em Proceedings of the 32nd ACM Multimedia}, 2024.

\bibitem[\protect\citeauthoryear{Chen \bgroup \em et al.\egroup }{2024g}]{chen2024talk}
Yuyan Chen, Yichen Yuan, Panjun Liu, Dayiheng Liu, Qinghao Guan, Mengfei Guo, Haiming Peng, Bang Liu, Zhixu Li, and Yanghua Xiao.
\newblock Talk funny! a large-scale humor response dataset with chain-of-humor interpretation.
\newblock In {\em Proceedings of the AAAI Conference on Artificial Intelligence}, volume~38, pages 17826--17834, 2024.

\bibitem[\protect\citeauthoryear{Chen \bgroup \em et al.\egroup }{2024h}]{chen2024temporalmed}
Yuyan Chen, Jin Zhao, Zhihao Wen, Zhixu Li, and Yanghua Xiao.
\newblock Temporalmed: Advancing medical dialogues with time-aware responses in large language models.
\newblock In {\em Proceedings of the 17th ACM International Conference on Web Search and Data Mining}, pages 116--124, 2024.

\bibitem[\protect\citeauthoryear{Ding \bgroup \em et al.\egroup }{2024}]{ding2024llava}
Zhicheng Ding, Panfeng Li, Qikai Yang, and Siyang Li.
\newblock Enhance image-to-image generation with llava-generated prompts.
\newblock In {\em 2024 5th International Conference on Information Science, Parallel and Distributed Systems (ISPDS)}, pages 77--81. IEEE, 2024.

\bibitem[\protect\citeauthoryear{Fu \bgroup \em et al.\egroup }{2024}]{fu2024detecting}
Zhe Fu, Kanlun Wang, Wangjiaxuan Xin, Lina Zhou, Shi Chen, Yaorong Ge, Daniel Janies, and Dongsong Zhang.
\newblock Detecting misinformation in multimedia content through cross-modal entity consistency: A dual learning approach.
\newblock 2024.

\bibitem[\protect\citeauthoryear{Huang \bgroup \em et al.\egroup }{2021}]{huang2021learning}
Junqin Huang, Linghe Kong, Jiejian Wu, Yutong Liu, Yuchen Li, and Zhe Wang.
\newblock Learning-based congestion control simulator for mobile internet education.
\newblock In {\em Proceedings of the 16th ACM Workshop on Mobility in the Evolving Internet Architecture}, pages 1--6, 2021.

\bibitem[\protect\citeauthoryear{J{\"{a}}rvelin and Kek{\"{a}}l{\"{a}}inen}{2017}]{DBLP:journals/sigir/JarvelinK17}
Kalervo J{\"{a}}rvelin and Jaana Kek{\"{a}}l{\"{a}}inen.
\newblock {IR} evaluation methods for retrieving highly relevant documents.
\newblock {\em {SIGIR} Forum}, 51(2):243--250, 2017.

\bibitem[\protect\citeauthoryear{Jiang \bgroup \em et al.\egroup }{2024}]{jiang2024multi}
Bowen Jiang, Yangxinyu Xie, Xiaomeng Wang, Weijie~J Su, Camillo~J Taylor, and Tanwi Mallick.
\newblock Multi-modal and multi-agent systems meet rationality: A survey.
\newblock {\em arXiv preprint arXiv:2406.00252}, 2024.

\bibitem[\protect\citeauthoryear{Jin \bgroup \em et al.\egroup }{2023}]{jin2022prototypical}
Yiqiao Jin, Xiting Wang, Yaru Hao, Yizhou Sun, and Xing Xie.
\newblock Prototypical fine-tuning: Towards robust performance under varying data sizes.
\newblock In {\em Proceedings of the AAAI Conference on Artificial Intelligence}, 2023.

\bibitem[\protect\citeauthoryear{Jin \bgroup \em et al.\egroup }{2024a}]{jin2024better}
Yiqiao Jin, Mohit Chandra, Gaurav Verma, Yibo Hu, Munmun De~Choudhury, and Srijan Kumar.
\newblock Better to ask in english: Cross-lingual evaluation of large language models for healthcare queries.
\newblock In {\em Web Conference}, pages 2627--2638, 2024.

\bibitem[\protect\citeauthoryear{Jin \bgroup \em et al.\egroup }{2024b}]{jin2024mm}
Yiqiao Jin, Minje Choi, Gaurav Verma, Jindong Wang, and Srijan Kumar.
\newblock Mm-soc: Benchmarking multimodal large language models in social media platforms.
\newblock In {\em ACL}, 2024.

\bibitem[\protect\citeauthoryear{Jin \bgroup \em et al.\egroup }{2024c}]{jin2024agentreview}
Yiqiao Jin, Qinlin Zhao, Yiyang Wang, Hao Chen, Kaijie Zhu, Yijia Xiao, and Jindong Wang.
\newblock Agentreview: Exploring peer review dynamics with llm agents.
\newblock In {\em EMNLP}, 2024.

\bibitem[\protect\citeauthoryear{Ke \bgroup \em et al.\egroup }{2017}]{lightgbm}
Guolin Ke, Qi~Meng, Thomas Finley, Taifeng Wang, Wei Chen, Weidong Ma, Qiwei Ye, and Tie{-}Yan Liu.
\newblock Lightgbm: {A} highly efficient gradient boosting decision tree.
\newblock In {\em Advances in Neural Information Processing Systems 30: Annual Conference on Neural Information Processing Systems}, pages 3146--3154, 2017.

\bibitem[\protect\citeauthoryear{Kingma and Welling}{2014}]{DBLP:journals/corr/KingmaW13}
Diederik~P. Kingma and Max Welling.
\newblock Auto-encoding variational bayes.
\newblock In {\em 2nd International Conference on Learning Representations}, 2014.

\bibitem[\protect\citeauthoryear{Li \bgroup \em et al.\egroup }{2020}]{DBLP:conf/icpr/LiL0R20}
Minghan Li, Xialei Liu, Joost van~de Weijer, and Bogdan~C. Raducanu.
\newblock Learning to rank for active learning: {A} listwise approach.
\newblock In {\em 25th International Conference on Pattern Recognition}, pages 5587--5594, 2020.

\bibitem[\protect\citeauthoryear{Li \bgroup \em et al.\egroup }{2022}]{li2022meta}
Yuchen Li, Haoyi Xiong, Linghe Kong, Rui Zhang, Dejing Dou, and Guihai Chen.
\newblock Meta hierarchical reinforced learning to rank for recommendation: A comprehensive study in moocs.
\newblock In {\em Joint European Conference on Machine Learning and Knowledge Discovery in Databases}, pages 302--317, 2022.

\bibitem[\protect\citeauthoryear{Li \bgroup \em et al.\egroup }{2023a}]{li2023mpgraf}
Yuchen Li, Haoyi Xiong, Linghe Kong, Zeyi Sun, Hongyang Chen, Shuaiqiang Wang, and Dawei Yin.
\newblock Mpgraf: a modular and pre-trained graphformer for learning to rank at web-scale.
\newblock In {\em 2023 IEEE International Conference on Data Mining (ICDM)}, pages 339--348. IEEE, 2023.

\bibitem[\protect\citeauthoryear{Li \bgroup \em et al.\egroup }{2023b}]{li2023s2phere}
Yuchen Li, Haoyi Xiong, Linghe Kong, Qingzhong Wang, Shuaiqiang Wang, Guihai Chen, and Dawei Yin.
\newblock S2phere: Semi-supervised pre-training for web search over heterogeneous learning to rank data.
\newblock In {\em Proceedings of the 29th ACM SIGKDD Conference on Knowledge Discovery and Data Mining}, pages 4437--4448, 2023.

\bibitem[\protect\citeauthoryear{Li \bgroup \em et al.\egroup }{2023c}]{li2023ltrgcn}
Yuchen Li, Haoyi Xiong, Linghe Kong, Shuaiqiang Wang, Zeyi Sun, Hongyang Chen, Guihai Chen, and Dawei Yin.
\newblock Ltrgcn: Large-scale graph convolutional networks-based learning to rank for web search.
\newblock In {\em Joint European Conference on Machine Learning and Knowledge Discovery in Databases}, pages 635--651. Springer, 2023.

\bibitem[\protect\citeauthoryear{Li \bgroup \em et al.\egroup }{2023d}]{li2023mhrr}
Yuchen Li, Haoyi Xiong, Linghe Kong, Rui Zhang, Fanqin Xu, Guihai Chen, and Minglu Li.
\newblock Mhrr: Moocs recommender service with meta hierarchical reinforced ranking.
\newblock {\em IEEE Transactions on Services Computing}, 2023.

\bibitem[\protect\citeauthoryear{Li \bgroup \em et al.\egroup }{2023e}]{li2023coltr}
Yuchen Li, Haoyi Xiong, Qingzhong Wang, Linghe Kong, Hao Liu, Haifang Li, Jiang Bian, Shuaiqiang Wang, Guihai Chen, Dejing Dou, et~al.
\newblock Coltr: Semi-supervised learning to rank with co-training and over-parameterization for web search.
\newblock {\em IEEE Transactions on Knowledge and Data Engineering}, 2023.

\bibitem[\protect\citeauthoryear{Li \bgroup \em et al.\egroup }{2024}]{li2024gs2p}
Yuchen Li, Haoyi Xiong, Linghe Kong, Jiang Bian, Shuaiqiang Wang, Guihai Chen, and Dawei Yin.
\newblock Gs2p: a generative pre-trained learning to rank model with over-parameterization for web-scale search.
\newblock {\em Machine Learning}, pages 1--19, 2024.

\bibitem[\protect\citeauthoryear{Liang \bgroup \em et al.\egroup }{2021}]{liang2021omnilytics}
Jiacheng Liang, Songze Li, Bochuan Cao, Wensi Jiang, and Chaoyang He.
\newblock Omnilytics: A blockchain-based secure data market for decentralized machine learning.
\newblock {\em arXiv preprint arXiv:2107.05252}, 2021.

\bibitem[\protect\citeauthoryear{Liang \bgroup \em et al.\egroup }{2024}]{liang2024model}
Jiacheng Liang, Ren Pang, Changjiang Li, and Ting Wang.
\newblock Model extraction attacks revisited.
\newblock In {\em Proceedings of the 19th ACM Asia Conference on Computer and Communications Security}, pages 1231--1245, 2024.

\bibitem[\protect\citeauthoryear{Liao \bgroup \em et al.\egroup }{2024}]{liao2024towards}
Yuan Liao, Jiang Bian, Yuhui Yun, Shuo Wang, Yubo Zhang, Jiaming Chu, Tao Wang, Kewei Li, Yuchen Li, Xuhong Li, et~al.
\newblock Towards automated data sciences with natural language and sagecopilot: Practices and lessons learned.
\newblock {\em arXiv preprint arXiv:2407.21040}, 2024.

\bibitem[\protect\citeauthoryear{Liu \bgroup \em et al.\egroup }{2024}]{liu2024robustifying}
Xiaoqun Liu, Jiacheng Liang, Muchao Ye, and Zhaohan Xi.
\newblock Robustifying safety-aligned large language models through clean data curation.
\newblock {\em arXiv preprint arXiv:2405.19358}, 2024.

\bibitem[\protect\citeauthoryear{Lu \bgroup \em et al.\egroup }{2024}]{lu2024cats}
Jiecheng Lu, Xu~Han, Yan Sun, and Shihao Yang.
\newblock Cats: Enhancing multivariate time series forecasting by constructing auxiliary time series as exogenous variables.
\newblock {\em arXiv preprint arXiv:2403.01673}, 2024.

\bibitem[\protect\citeauthoryear{Lyu \bgroup \em et al.\egroup }{2020}]{lyu2020movement}
Zhonghao Lyu, Chenhao Ren, and Ling Qiu.
\newblock Movement and communication co-design in multi-uav enabled wireless systems via drl.
\newblock In {\em 2020 IEEE 6th International Conference on Computer and Communications (ICCC)}, pages 220--226. IEEE, 2020.

\bibitem[\protect\citeauthoryear{Lyu \bgroup \em et al.\egroup }{2022a}]{lyu2022joint1}
Zhonghao Lyu, Guangxu Zhu, and Jie Xu.
\newblock Joint maneuver and beamforming design for uav-enabled integrated sensing and communication.
\newblock {\em IEEE Transactions on Wireless Communications}, 22(4):2424--2440, 2022.

\bibitem[\protect\citeauthoryear{Lyu \bgroup \em et al.\egroup }{2022b}]{lyu2022joint}
Zhonghao Lyu, Guangxu Zhu, and Jie Xu.
\newblock Joint trajectory and beamforming design for uav-enabled integrated sensing and communication.
\newblock In {\em ICC 2022-IEEE International Conference on Communications}, pages 1593--1598. IEEE, 2022.

\bibitem[\protect\citeauthoryear{Lyu \bgroup \em et al.\egroup }{2023}]{lyu2023semantic}
Zhonghao Lyu, Guangxu Zhu, Jie Xu, Bo~Ai, and Shuguang Cui.
\newblock Semantic communications for joint image recovery and classification.
\newblock In {\em 2023 IEEE Globecom Workshops (GC Wkshps)}, pages 1579--1584. IEEE, 2023.

\bibitem[\protect\citeauthoryear{Lyu \bgroup \em et al.\egroup }{2024a}]{lyu2024rethinking}
Zhonghao Lyu, Yuchen Li, Guangxu Zhu, Jie Xu, H~Vincent Poor, and Shuguang Cui.
\newblock Rethinking resource management in edge learning: A joint pre-training and fine-tuning design paradigm.
\newblock {\em arXiv preprint arXiv:2404.00836}, 2024.

\bibitem[\protect\citeauthoryear{Lyu \bgroup \em et al.\egroup }{2024b}]{lyu2024semantic}
Zhonghao Lyu, Guangxu Zhu, Jie Xu, Bo~Ai, and Shuguang Cui.
\newblock Semantic communications for image recovery and classification via deep joint source and channel coding.
\newblock {\em IEEE Transactions on Wireless Communications}, 2024.

\bibitem[\protect\citeauthoryear{Ni \bgroup \em et al.\egroup }{2024}]{ni2024timeseries}
Haowei Ni, Shuchen Meng, Xieming Geng, Panfeng Li, Zhuoying Li, Xupeng Chen, Xiaotong Wang, and Shiyao Zhang.
\newblock Time series modeling for heart rate prediction: From arima to transformers.
\newblock {\em arXiv preprint arXiv:2406.12199}, 2024.

\bibitem[\protect\citeauthoryear{Peng \bgroup \em et al.\egroup }{2023}]{peng2023clgt}
Tianhao Peng, Yu~Liang, Wenjun Wu, Jian Ren, Zhao Pengrui, and Yanjun Pu.
\newblock Clgt: A graph transformer for student performance prediction in collaborative learning.
\newblock In {\em Proceedings of the AAAI conference on artificial intelligence}, volume~37, pages 15947--15954, 2023.

\bibitem[\protect\citeauthoryear{Peng \bgroup \em et al.\egroup }{2024}]{peng2024graphrare}
Tianhao Peng, Wenjun Wu, Haitao Yuan, Zhifeng Bao, Zhao Pengru, Xin Yu, Xuetao Lin, Yu~Liang, and Yanjun Pu.
\newblock Graphrare: Reinforcement learning enhanced graph neural network with relative entropy.
\newblock In {\em 2024 IEEE 40th International Conference on Data Engineering (ICDE)}, pages 2489--2502. IEEE, 2024.

\bibitem[\protect\citeauthoryear{Pobrotyn and Bia{\l}obrzeski}{2021}]{DBLP:journals/corr/abs-2102-07831}
Przemys{\l}aw Pobrotyn and Rados{\l}aw Bia{\l}obrzeski.
\newblock Neuralndcg: Direct optimisation of a ranking metric via differentiable relaxation of sorting.
\newblock {\em arXiv preprint arXiv:2102.07831}, 2021.

\bibitem[\protect\citeauthoryear{Pobrotyn \bgroup \em et al.\egroup }{2020}]{context}
Przemys{\l}aw Pobrotyn, Tomasz Bartczak, Miko{\l}aj Synowiec, Rados{\l}aw Bia{\l}obrzeski, and Jaros{\l}aw Bojar.
\newblock Context-aware learning to rank with self-attention.
\newblock {\em arXiv preprint arXiv:2005.10084}, 2020.

\bibitem[\protect\citeauthoryear{Qin and Liu}{2013}]{DBLP:journals/corr/QinL13}
Tao Qin and Tie-Yan Liu.
\newblock Introducing letor 4.0 datasets.
\newblock {\em arXiv preprint arXiv:1306.2597}, 2013.

\bibitem[\protect\citeauthoryear{Qin \bgroup \em et al.\egroup }{2010}]{DBLP:journals/ir/QinLL10}
Tao Qin, Tie{-}Yan Liu, and Hang Li.
\newblock A general approximation framework for direct optimization of information retrieval measures.
\newblock {\em Inf. Retr.}, 13(4):375--397, 2010.

\bibitem[\protect\citeauthoryear{Rahimi and Recht}{2007}]{DBLP:conf/nips/RahimiR07}
Ali Rahimi and Benjamin Recht.
\newblock Random features for large-scale kernel machines.
\newblock In {\em Advances in Neural Information Processing Systems 20, Proceedings of the Twenty-First Annual Conference on Neural Information Processing Systems}, pages 1177--1184, 2007.

\bibitem[\protect\citeauthoryear{Shangguan \bgroup \em et al.\egroup }{2021}]{shangguan2021trend}
Zhongkai Shangguan, Zihe Zheng, and Lei Lin.
\newblock Trend and thoughts: Understanding climate change concern using machine learning and social media data.
\newblock {\em arXiv preprint arXiv:2111.14929}, 2021.

\bibitem[\protect\citeauthoryear{Song \bgroup \em et al.\egroup }{2023}]{song_going_2023}
Yukun Song, Parth Arora, Rajandeep Singh, Srikanth~T. Varadharajan, Malcolm Haynes, and Thad Starner.
\newblock Going blank comfortably: Positioning monocular head-worn displays when they are inactive.
\newblock In {\em Proceedings of the 2023 ACM International Symposium on Wearable Computers}, ISWC '23, page 114–118, New York, NY, USA, October 2023. Association for Computing Machinery.

\bibitem[\protect\citeauthoryear{Szummer and Yilmaz}{2011}]{DBLP:conf/cikm/SzummerY11}
Martin Szummer and Emine Yilmaz.
\newblock Semi-supervised learning to rank with preference regularization.
\newblock In {\em Proceedings of the 20th {ACM} Conference on Information and Knowledge Management}, pages 269--278, 2011.

\bibitem[\protect\citeauthoryear{Tran \bgroup \em et al.\egroup }{2017}]{DBLP:conf/cvpr/Tran0ZJ17}
Luan Tran, Xiaoming Liu, Jiayu Zhou, and Rong Jin.
\newblock Missing modalities imputation via cascaded residual autoencoder.
\newblock In {\em 2017 {IEEE} Conference on Computer Vision and Pattern Recognition}, pages 4971--4980, 2017.

\bibitem[\protect\citeauthoryear{Vaswani \bgroup \em et al.\egroup }{2017}]{DBLP:conf/nips/VaswaniSPUJGKP17}
Ashish Vaswani, Noam Shazeer, Niki Parmar, Jakob Uszkoreit, Llion Jones, Aidan~N. Gomez, Lukasz Kaiser, and Illia Polosukhin.
\newblock Attention is all you need.
\newblock In {\em Advances in Neural Information Processing Systems 30: Annual Conference on Neural Information Processing Systems}, pages 5998--6008, 2017.

\bibitem[\protect\citeauthoryear{Vincent \bgroup \em et al.\egroup }{2010}]{DBLP:journals/jmlr/VincentLLBM10}
Pascal Vincent, Hugo Larochelle, Isabelle Lajoie, Yoshua Bengio, and Pierre{-}Antoine Manzagol.
\newblock Stacked denoising autoencoders: Learning useful representations in a deep network with a local denoising criterion.
\newblock {\em J. Mach. Learn. Res.}, 11:3371--3408, 2010.

\bibitem[\protect\citeauthoryear{Wang \bgroup \em et al.\egroup }{2021}]{DBLP:conf/www/WangSCJLHC21}
Ruoxi Wang, Rakesh Shivanna, Derek~Zhiyuan Cheng, Sagar Jain, Dong Lin, Lichan Hong, and Ed~H. Chi.
\newblock {DCN} {V2:} improved deep {\&} cross network and practical lessons for web-scale learning to rank systems.
\newblock In {\em {WWW} '21: The Web Conference}, pages 1785--1797, 2021.

\bibitem[\protect\citeauthoryear{Wang \bgroup \em et al.\egroup }{2023a}]{wang2023st}
Zepu Wang, Yuqi Nie, Peng Sun, Nam~H Nguyen, John Mulvey, and H~Vincent Poor.
\newblock St-mlp: A cascaded spatio-temporal linear framework with channel-independence strategy for traffic forecasting.
\newblock {\em arXiv preprint arXiv:2308.07496}, 2023.

\bibitem[\protect\citeauthoryear{Wang \bgroup \em et al.\egroup }{2023b}]{wang2023novel}
Zepu Wang, Peng Sun, Yulin Hu, and Azzedine Boukerche.
\newblock A novel hybrid method for achieving accurate and timeliness vehicular traffic flow prediction in road networks.
\newblock {\em Computer Communications}, 209:378--386, 2023.

\bibitem[\protect\citeauthoryear{Wang \bgroup \em et al.\egroup }{2023c}]{wang2023sst}
Zepu Wang, Yifei Sun, Zhiyu Lei, Xincheng Zhu, and Peng Sun.
\newblock Sst: A simplified swin transformer-based model for taxi destination prediction based on existing trajectory.
\newblock In {\em 2023 IEEE 26th International Conference on Intelligent Transportation Systems (ITSC)}, pages 1404--1409. IEEE, 2023.

\bibitem[\protect\citeauthoryear{Wang \bgroup \em et al.\egroup }{2024a}]{wang2024multi}
Ning Wang, Jiang Bian, Yuchen Li, Xuhong Li, Shahid Mumtaz, Linghe Kong, and Haoyi Xiong.
\newblock Multi-purpose rna language modelling with motif-aware pretraining and type-guided fine-tuning.
\newblock {\em Nature Machine Intelligence}, pages 1--10, 2024.

\bibitem[\protect\citeauthoryear{Wang \bgroup \em et al.\egroup }{2024b}]{wang2024soft}
Qunbo Wang, Ruyi Ji, Tianhao Peng, Wenjun Wu, Zechao Li, and Jing Liu.
\newblock Soft knowledge prompt: Help external knowledge become a better teacher to instruct llm in knowledge-based vqa.
\newblock In {\em Proceedings of the 62nd Annual Meeting of the Association for Computational Linguistics (Volume 1: Long Papers)}, pages 6132--6143, 2024.

\bibitem[\protect\citeauthoryear{Wang \bgroup \em et al.\egroup }{2024c}]{wang2024uncertainty}
Zepu Wang, Xiaobo Ma, Huajie Yang, Weimin Lvu, Peng Sun, and Sharath~Chandra Guntuku.
\newblock Uncertainty-aware crime prediction with spatial temporal multivariate graph neural networks.
\newblock {\em arXiv preprint arXiv:2408.04193}, 2024.

\bibitem[\protect\citeauthoryear{Weng and Wu}{2024a}]{Weng202404}
Yijie Weng and Jianhao Wu.
\newblock Big data and machine learning in defence.
\newblock {\em International Journal of Computer Science and Information Technology}, 16(2), 2024.

\bibitem[\protect\citeauthoryear{Weng and Wu}{2024b}]{weng2024leveraging}
Yijie Weng and Jianhao Wu.
\newblock Leveraging artificial intelligence to enhance data security and combat cyber attacks.
\newblock {\em Journal of Artificial Intelligence General science (JAIGS) ISSN: 3006-4023}, 5(1):392--399, 2024.

\bibitem[\protect\citeauthoryear{Werner}{2022}]{DBLP:journals/ml/Werner22}
Tino Werner.
\newblock A review on instance ranking problems in statistical learning.
\newblock {\em Mach. Learn.}, 111(2):415--463, 2022.

\bibitem[\protect\citeauthoryear{Xia \bgroup \em et al.\egroup }{2008}]{DBLP:conf/icml/XiaLWZL08}
Fen Xia, Tie{-}Yan Liu, Jue Wang, Wensheng Zhang, and Hang Li.
\newblock Listwise approach to learning to rank: theory and algorithm.
\newblock In {\em Machine Learning, Proceedings of the Twenty-Fifth International Conference}, pages 1192--1199, 2008.

\bibitem[\protect\citeauthoryear{Xie \bgroup \em et al.\egroup }{2024}]{xie2024wildfiregpt}
Yangxinyu Xie, Tanwi Mallick, Joshua~David Bergerson, John~K Hutchison, Duane~R Verner, Jordan Branham, M~Ross Alexander, Robert~B Ross, Yan Feng, Leslie-Anne Levy, et~al.
\newblock Wildfiregpt: Tailored large language model for wildfire analysis.
\newblock {\em arXiv preprint arXiv:2402.07877}, 2024.

\bibitem[\protect\citeauthoryear{Xin \bgroup \em et al.\egroup }{2024}]{xin2024let}
Wangjiaxuan Xin, Kanlun Wang, Zhe Fu, and Lina Zhou.
\newblock Let community rules be reflected in online content moderation.
\newblock {\em arXiv preprint arXiv:2408.12035}, 2024.

\bibitem[\protect\citeauthoryear{Xiong \bgroup \em et al.\egroup }{2024a}]{xiong2024search}
Haoyi Xiong, Jiang Bian, Yuchen Li, Xuhong Li, Mengnan Du, Shuaiqiang Wang, Dawei Yin, and Sumi Helal.
\newblock When search engine services meet large language models: Visions and challenges.
\newblock {\em IEEE Transactions on Services Computing}, 2024.

\bibitem[\protect\citeauthoryear{Xiong \bgroup \em et al.\egroup }{2024b}]{xiong2024towards}
Haoyi Xiong, Xiaofei Zhang, Jiamin Chen, Xinhao Sun, Yuchen Li, Zeyi Sun, Mengnan Du, et~al.
\newblock Towards explainable artificial intelligence (xai): A data mining perspective.
\newblock {\em arXiv preprint arXiv:2401.04374}, 2024.

\bibitem[\protect\citeauthoryear{Yang and Ying}{2023}]{DBLP:journals/csur/YangY23}
Tianbao Yang and Yiming Ying.
\newblock {AUC} maximization in the era of big data and {AI:} {A} survey.
\newblock {\em {ACM} Comput. Surv.}, 55(8):172:1--172:37, 2023.

\bibitem[\protect\citeauthoryear{Yu \bgroup \em et al.\egroup }{2018}]{yu2018adaptively}
Chao Yu, Dongxu Wang, Tianpei Yang, Wenxuan Zhu, Yuchen Li, Hongwei Ge, and Jiankang Ren.
\newblock Adaptively shaping reinforcement learning agents via human reward.
\newblock In {\em PRICAI 2018: Trends in Artificial Intelligence: 15th Pacific Rim International Conference on Artificial Intelligence, Nanjing, China, August 28--31, 2018, Proceedings, Part I 15}, pages 85--97. Springer, 2018.

\bibitem[\protect\citeauthoryear{Zhang \bgroup \em et al.\egroup }{2016}]{DBLP:journals/www/ZhangHL16}
Xin Zhang, Ben He, and Tiejian Luo.
\newblock Training query filtering for semi-supervised learning to rank with pseudo labels.
\newblock {\em World Wide Web}, 19(5):833--864, 2016.

\bibitem[\protect\citeauthoryear{Zhao \bgroup \em et al.\egroup }{2010}]{DBLP:conf/coling/ZhaoWL10}
Shiqi Zhao, Haifeng Wang, and Ting Liu.
\newblock Paraphrasing with search engine query logs.
\newblock In {\em {COLING} 2010, 23rd International Conference on Computational Linguistics, Proceedings of the Conference}, pages 1317--1325, 2010.

\bibitem[\protect\citeauthoryear{Zhao \bgroup \em et al.\egroup }{2011}]{zhao2011automatically}
Shiqi Zhao, Haifeng Wang, Chao Li, Ting Liu, and Yi~Guan.
\newblock Automatically generating questions from queries for community-based question answering.
\newblock In {\em Proceedings of 5th international joint conference on natural language processing}, pages 929--937, 2011.

\bibitem[\protect\citeauthoryear{Zheng \bgroup \em et al.\egroup }{2021}]{zheng2021makes}
Zihe Zheng, Zhongkai Shangguan, and Jiebo Luo.
\newblock What makes a turing award winner?
\newblock In {\em Social, Cultural, and Behavioral Modeling: 14th International Conference, SBP-BRiMS 2021, Virtual Event, July 6--9, 2021, Proceedings 14}, pages 310--320. Springer, 2021.

\bibitem[\protect\citeauthoryear{Zhou \bgroup \em et al.\egroup }{2024}]{zhou2024pass}
Qihua Zhou, Song Guo, Jun Pan, Jiacheng Liang, Jingcai Guo, Zhenda Xu, and Jingren Zhou.
\newblock Pass: Patch automatic skip scheme for efficient on-device video perception.
\newblock {\em IEEE Transactions on Pattern Analysis and Machine Intelligence}, 2024.

\bibitem[\protect\citeauthoryear{Zhou}{2024a}]{zhou2024application}
Qiqin Zhou.
\newblock Application of black-litterman bayesian in statistical arbitrage.
\newblock {\em arXiv preprint arXiv:2406.06706}, 2024.

\bibitem[\protect\citeauthoryear{Zhou}{2024b}]{zhou2024explainable}
Qiqin Zhou.
\newblock Explainable ai in request-for-quote.
\newblock {\em arXiv preprint arXiv:2407.15038}, 2024.

\bibitem[\protect\citeauthoryear{Zhou}{2024c}]{zhou2024portfolio}
Qiqin Zhou.
\newblock Portfolio optimization with robust covariance and conditional value-at-risk constraints.
\newblock {\em arXiv preprint arXiv:2406.00610}, 2024.

\bibitem[\protect\citeauthoryear{Zhu \bgroup \em et al.\egroup }{2023}]{zhu2023pushing}
Guangxu Zhu, Zhonghao Lyu, Xiang Jiao, Peixi Liu, Mingzhe Chen, Jie Xu, Shuguang Cui, and Ping Zhang.
\newblock Pushing ai to wireless network edge: An overview on integrated sensing, communication, and computation towards 6g.
\newblock {\em Science China Information Sciences}, 66(3):130301, 2023.

\end{thebibliography}

\end{document}